\documentclass[12pt]{iopart}
\usepackage{graphics}
\usepackage{cite}
\begin{document}
\def\Up{\Uparrow}
\def\Dn{\Downarrow}
\jl{3}
\title
[Field-Induced gap due to four-spin exchange in a spin ladder]
{Field-Induced gap due to four-spin exchange in a spin ladder}
\author{Atsushi Nakasu\dag, 
Keisuke Totsuka\ddag\footnote[4]
{Present address: Department of Physics, Aoyama Gakuin University, 
Setagaya-ku, Tokyo 157-0071, Japan}, 
Yasumasa Hasegawa\dag, Kiyomi Okamoto\S\footnote[5]
{To whom correspondence should be addressed} and
T\^oru Sakai\dag\footnote[6]
{Present address: Tokyo Metropolitan Institute of Technology, 
Hino-shi, Tokyo 191-0065, Japan}}

\address{\dag\ Faculty of Science, 
         Himeji Institute of Technology, 
         Ako-gun, Hyogo 678-1297, Japan}

\address{\ddag\ Department of Physics,
         Kyushu University, 
         Higashi-ku, Fukuoka 812-8581, Japan}

\address{\S\ Department of Physics,
         Tokyo Institute of Technology,
         Meguro-ku, Tokyo 152-8551, Japan}

\begin{abstract}
The effect of the four-spin cyclic exchange interaction at each plaquette in 
the $S=1/2$ two-leg spin ladder is investigated at $T=0$,
especially focusing on the field-induced gap. 
The strong rung coupling approximation suggests that 
it yields a plateau at half of the saturation moment ($m=1/2$) in the 
magnetization curve, which corresponds to a field-induced 
spin gap with a spontaneous breaking of the translational symmetry. 
A precise phase diagram at $m=1/2$ is also presented based on the 
level spectroscopy analysis of the numerical data obtained by
Lanczos method. 
The boundary between the gapless and plateau phases is 
confirmed to be of the Kosterlitz-Thouless (KT) universality class. 
\end{abstract}
\maketitle
\section{Introduction}

The multiple-spin exchange interaction 
has attracted a lot of interest in the condensed matter physics. 
Such many-body exchange
interactions are realized in the two-dimensional (2D)
solid $^3$He \cite{he1,he2} and the 2D
Wigner solid of electrons formed
in a silicon inversion layer \cite{wigner},
as well as the bcc $^3$He \cite{he3}.
Recently the four-spin cyclic exchange interaction, 
called {\it ring exchange}, has been revealed to be 
important even in the strongly correlated electron systems 
like the high-temperature cuprate superconductors.   
The analysis on the low-lying excitation spectrum of the $d$-$p$ model 
indicated that the ring exchange should be taken into account 
in the simplified spin Hamiltonian describing the CuO$_2$ 
plane \cite{schmidt}. 
The study based on the Heisenberg Hamiltonian also revealed 
an evidence of the ring exchange in the 
Raman scattering spectrum \cite{honda1}.  
In fact such a four-spin interaction was derived from the 
fourth-order perturbation expansion of the square lattice 
Hubbard Hamiltonian 
with respect to $t/U$ near half-filling \cite{mt}. 
The neutron scattering experiment also suggested that 
the effect of the ring exchange appeared in the spin wave 
excitation spectrum of La$_2$CuO$_4$ \cite{neutron}.  
 
Recently the ring exchange has been supposed to be important 
also in the spin ladder systems. 
The significant difference between observed leg and rung bi-linear exchange 
coupling constants ($J_{\rm leg} \sim 2 J_{\rm rung}$) 
of Sr$_{14}$Cu$_{24}$O$_{41}$ was explained,  
assuming the existence of the ring exchange with the amplitude of 
about 14\% of $J_{\rm rung}$ \cite{brehmer}.   
A four-spin exchange interaction described by a product of two-spin
exchanges in a spin ladder
was investigated by a field theoretical approach \cite{tsvelik},
where the possibility of a different type of massive phase
from the Haldane phase is indicated \cite{haldane1, haldane2}.
in the nonmagnetic ground state. 
The recent density matrix renormalization group study suggested 
that increasing ring exchange constant $J_4$ brings about a quantum 
phase transition about $J_4 \sim 0.3 J_{\rm rung}$ for 
$J_{\rm leg}=J_{\rm rung}$ \cite{honda2}.  
However, the feature of the large-$J_4$ phase is still an open problem. 
We note that there exist integrable spin ladder models with four-spin
exchange interactions \cite{wang, links}.
But these models are somewhat differnt from our model.

%
%  Fig. 1
%
\begin{figure}[h]
   \begin{center}
         \scalebox{0.4}[0.4]{\includegraphics{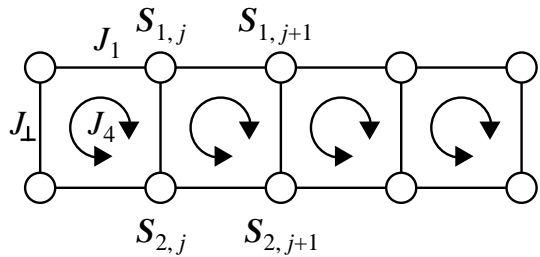}}
   \end{center}
\caption{Illustration of a two-leg spin ladder with four-spin exchange interaction
         at every plaquette.}
\label{model}
\end{figure}

As another interesting phenomenon caused by the ring exchange, 
a field-induced spin gap \cite{oshikawa}, 
which should be observed as a plateau 
in the magnetization curve,  was proposed in an $S=1/2$ 
antiferromagnet on the triangular lattice \cite{kubo}.  
It was verified by the exact diagonalization \cite{misguich}.
In the spin ladder system, 
besides the original spin gap \cite{hida,dagotto,troyer}, 
the magnetization plateau was also revealed to appear  
in the presence of an additional leg \cite{cabra}, 
a bond alternation \cite{cabra2,okamoto-okazaki-sakai-2} 
or some frustrated interactions
\cite{tonegawa,tone-oka-kabu,totsuka,mila,okazaki1,okazaki2,okamoto-okazaki-sakai-1}.  
In the previous work \cite{sh} 
by two of the present authors (TS and YH), 
such a field induced gap due to the ring exchange in the 
spin ladder was investigated. 
In that work, 
the phenomenological renormalization study \cite{phenomenological} 
suggested that a plateau 
appeared at half of the saturation magnetization in the 
spin ladder for a realistic parameter $J_4 > (0.05 \pm 0.04) 
J_{\rm rung}$ in the case of $J_{\rm leg}=J_{\rm rung}$. 
However, the critical point $J_{\rm 4c}$ estimated by 
the one-magnon analysis did not agree well 
with that by the two-magnon one. 
The critical index at $J_{\rm 4c}$ was also different from  
that of the Kosterlitz-Thouless transition \cite{kt} 
predicted by the behavior of the energy gap.  
Thus some better studies are necessary to determine the critical 
point $J_{\rm 4c}$ and the universality class for 
the transition between the plateau and gapless states
due to the ring exchange in the spin ladder.  
In the present paper, 
the perturbation expansion from the strong rung coupling limit 
is performed to clarify the mechanism of the plateau formation
and the quantum critical behavior around the phase boundary. 
Also, a precise phase diagram is presented,  
using the size scaling analysis based on the conformal 
field theory \cite{cft1, cft2, cft3} 
and the recently-developed level spectroscopy
method \cite{level-1,level-2,level-3,nomura}. 
The parameter space is also more generalized than that of the previous 
work. 
 
\section{Spin Ladder with Ring Exchange}

We consider the $S=1/2$ uniform antiferromagnetic spin ladder
with the four-spin cyclic exchange at every plaquette,
as shown in figure 1.
The Hamiltonian of our model is
\begin{eqnarray}
\label{ham}
&{\cal H}&={\cal H}_0+{\cal H}_Z, \nonumber \\
&{\cal H}_0&=\sum_{j=1}^L(J_1{\bi S}_{1,j}\cdot {\bi S}_{1,j+1}
               +J_1{\bi S}_{2,j}\cdot {\bi S}_{2,j+1}
               +J_{\perp}{\bi S}_{1,j}\cdot {\bi S}_{2,j})
 \nonumber \\
&&   +  J_4 \sum_{j=1}^L (P_{4,j}+P^{-1}_{4,j})
, \\
&{\cal H}_Z& =-H\sum_{j=1}^L (S_{1,j}^z+S_{2,j}^z), \nonumber
\end{eqnarray}
where $J_1$ and $J_{\perp}$ are the bi-linear leg and rung exchange constants,
respectively. We set $J_{\perp}=1$ throughout this paper. 
$P_{4,j}$ is the cyclic permutation operator which
exchanges the four spins around the $j$-th plaquette as
${\bi S}_{1,j}\rightarrow {\bi S}_{1,j+1}\rightarrow
{\bi S}_{2,j+1}\rightarrow {\bi S}_{2,j}\rightarrow {\bi S}_{1,j}$,
\begin{equation}
    P_j \left| \matrix{a &b \cr
                       d &c }
        \right\rangle
     =  \left| \matrix{d &a \cr
                       c &b}
        \right\rangle,~~~~~             
    P_j^{-1} \left| \matrix{a &b \cr
                       d &c }
        \right\rangle
     =  \left| \matrix{b &c \cr
                       a &d}
        \right\rangle              
\end{equation}
where the $(1,1)$ component denotes the spin state of ${\bi S}_{1,j}$
and the $(2,2)$ component that of ${\bi S}_{2,j+1}$.
The strength of the four-spin ring exchange $J_4$ is assumed to be
positive, as it is in the Cu oxides.  
The applied magnetic field is denoted by $H$ .
The magnetization of the bulk system is defined as $m=M/L$, 
where $M\equiv \langle \sum _j(S_{1,j}^z+S_{2,j}^z)\rangle$. 
We focus on a possible plateau at half of the saturation ($m=1/2$)   
in the magnetization curve at $T=0$. 

\section{Strong rung coupling approximation}

The condition of the quantization of the magnetization \cite{oshikawa} 
predicted that $Q(S-m)=$integer is necessary for the appearance of 
the plateau at $m$, where $Q$ is the period of the ground state 
and $S$ is the total spin per unit cell. 
Thus the plateau at $m=1/2$ should require a spontaneous breakdown 
of the translational symmetry which results in the period of the 
two unit cells like the N\'eel state.  

In order to investigate the possibility and the mechanism of 
the plateau at $m=1/2$ in the spin ladder with the ring exchange, 
we consider the strong rung coupling limit 
($J_{\perp} \gg J_1, J_4$) \cite{totsuka,mila} at first. 
The system is separated into isolated rung dimers when $J_1 = J_4 =0$. 
In this case, at $m=1/2$, half of the rung pairs are triplet with $S^z =1$,
and the remaining half are singlet.
Since other possible selections have higher energies, we may 
restrict ourselves to these two states for each rung dimer 
and represent them by states
$| \Uparrow \rangle$ and $| \Downarrow \rangle$
of pseudo-spin $\tilde{\bi S}$
\begin{equation}
    \left| \matrix{\uparrow \cr \uparrow}\right\rangle 
    \Rightarrow | \Uparrow \rangle,~~~~~
    {1 \over \sqrt{2}}
    \left(
    \left| \matrix{\uparrow \cr \downarrow}\right\rangle
     - \left| \matrix{\downarrow \cr \uparrow}\right\rangle
    \right)
    \Rightarrow | \Downarrow \rangle    \; . 
\end{equation}
When $J_1 = J_4 =0$, 
each pseudospin $\tilde{\bi S}$ can take either 
$| \Uparrow \rangle$ or $| \Downarrow \rangle$ completely freely,
which means high degeneracy of the $m=1/2$ ground states. 
The introduction of weak $J_{1}$ and $J_{4}$ resolves the degeneracy.   
The effect of a weak inter-dimer coupling can be taken into account 
by the perturbation expansion with respect to $J_{1}$ and $J_{4}$, 
resulting in the following matrix elements by straightforward calculation
\begin{equation}\fl
    \bordermatrix{
    &| \Up_j \Up_{j+1} \rangle
    &| \Up_j \Dn_{j+1} \rangle
    &| \Dn_j \Up_{j+1} \rangle
    &| \Dn_j \Dn_{j+1} \rangle \cr
    \langle \Up_j \Up_{j+1} | &-(1/4)J_1+2J_4 &0 &0 &0 \cr
    \langle \Up_j \Dn_{j+1} | &0 &-(1/4)J_1-J_4 &(1/2)J_1+J_4  &0 \cr
    \langle \Dn_j \Up_{j+1} | &0 &(1/2)J_1+J_4  &-(1/4)J_1-J_4 &0 \cr
    \langle \Dn_j \Dn_{j+1} | &0 &0 &0 &(1/4)J_1+J_4}
    \label{eq:matrix}
\end{equation}
By comparing equation (\ref{eq:matrix}) with an $XXZ$-type Hamiltonian 
including an effective magnetic field $\tilde H$
\begin{equation}\fl
    \bordermatrix{
    &| \Up_j \Up_{j+1} \rangle
    &| \Up_j \Dn_{j+1} \rangle
    &| \Dn_j \Up_{j+1} \rangle
    &| \Dn_j \Dn_{j+1} \rangle \cr
    \langle \Up_j \Up_{j+1} | &(1/4)\tilde J^z-\tilde H+a &0 &0 &0 \cr
    \langle \Up_j \Dn_{j+1} | &0 &-(1/4)\tilde J^z+a &(1/2)\tilde J^{xy} &0 \cr
    \langle \Dn_j \Up_{j+1} | &0 &(1/2)\tilde J^{xy}  &-(1/4)\tilde J^z+a &0 \cr
    \langle \Dn_j \Dn_{j+1} | &0 &0 &0 &(1/4)\tilde J^z+\tilde H+a} \; ,
    \label{eq:matrix2}
\end{equation}
we obtain the effective Hamiltonian for $\tilde{\bi S}$
\begin{equation}
\label{eham}
   {\tilde{\cal H}}=
   \sum_j^L \left\{ \tilde J^{xy} (\tilde{S}_j^x\tilde{S}_{j+1}^x
   + \tilde{S}_j^y\tilde{S}_{j+1}^y)
   + \tilde J^z \tilde{S}_j^z\tilde{S}_{j+1}^z \right\}
   +{\tilde{\cal H}}_{\rm Z} \; 
\end{equation}
with $\tilde{J}^{xy}=J_{1}+2J_{4}$ and 
$\tilde{J}^{z}=J_{1}/2+5J_{4}$.   

The effective Zeeman term ${\tilde{\cal H}}_Z$   
and the effective $XXZ$ anisotropy are given respectively by  
\begin{equation}
    \tilde{\cal H}_{\rm Z}
    = \tilde H \sum_j^L \tilde S_j^z~~~~~~~
    \tilde H = {J_1 \over 4} - {J_4 \over 2}
\end{equation}
and  
\begin{equation}
   \lambda_{\rm eff} \equiv  {\tilde J^z \over \tilde J^{xy}}
   = {{J_1/2+5J_4}\over {J_1+2J_4}}  \; .
\end{equation}
Thus the problem of magnetized states with $m=1/2$ of the {\em original} 
system 
is mapped onto that of nonmagnetic states of the Hamiltonian (\ref{eham}).
It is well-known for the $\tilde S=1/2$ $XXZ$ chain that 
the ground state has the N\'eel order for $\lambda _{\rm eff}>1$  
and magnetic/non-magnetic excitations are gapped.   
The ordered phase is separated from the gapless one by 
a critical point $\lambda _{\rm eff}=1$;      
the trasition at the critical point is of 
the Kosterlitz-Thouless type with $\eta = \eta^z = 1$, 
where $\eta$ and $\eta^z$ are critical indices defined by 
\begin{equation}
\langle S^x_0 S^x_r\rangle \sim (-1)^r r^{-\eta}~~~~ 
\mbox{ and }  
\langle S^z_0 S^z_r\rangle \sim (-1)^r r^{-\eta^z}~~~~ 
(r \rightarrow \infty) 
\end{equation}
respectively.  
Turning back to the original system, this implies that 
the original ladder (\ref{ham}) has a magnetization plateau 
at $m=1/2$ under the condition 
\begin{eqnarray}
\label{condition1}
J_4 > {1\over 6}J_1. 
\end{eqnarray}
The plateau state is N\'eel ordered in the language of the pseudo-spin
$\tilde{\bi S}$, as shown in figure 2.
%
%  Fig. 2
%
\begin{figure}[h]
   \begin{center}
         \scalebox{0.6}[0.6]{\includegraphics{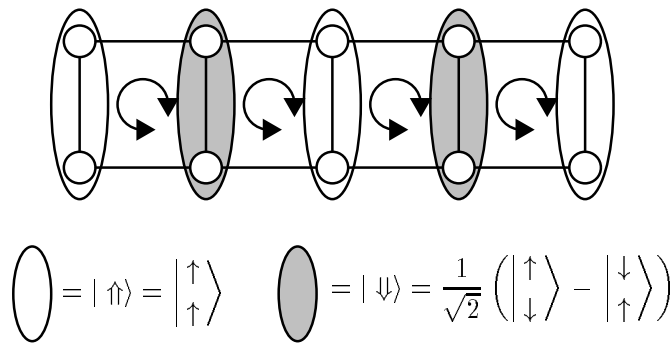}}
   \end{center}
\caption{Physical picture of the $m=1/2$ plateau state.
Open ellipses represent $|\Uparrow\rangle$ and shadowed ellipses 
$|\Downarrow\rangle$ in the language of the pseudo-spin $\tilde{\bi S}$.}
\label{model}
\end{figure}

\section{Numerical approach}

The phase boudnary $J_4 = (1/6)J_1$ obtained in the previous section
is valid for $J_1, J_4 \ll 1$.
To obtain a phase diagram for wider range of the parameters,
we have performed the numerical diagonalization of the
original Hamiltonian (\ref{ham}) by the Lanczos method.

The most powerful method at the present stage to determine
the KT phase boundary is the level spectroscopy (LS), 
which was developed by one of the present authors (KO) and Nomura 
\cite{level-1,level-2,level-3,nomura}. 
In the LS analysis the critical point is determined from the level cross 
between the two relevant low-lying excitation gaps with a 
common scaling dimension, that is, the same dominant size 
correction including the logarithmic one. 
In the present case, 
we use the following three gaps  
\begin{eqnarray}
  \label{gap1}
  \Delta_1 &=& {1\over 2} [E_{\pi}(L,M+1)+E_{\pi}(L,M-1)-2E_0(L,M)] \\
  \Delta_{\pi}&=& E_{\pi}(L,M)-E_0(L,M) \\
  \Delta_{\pi}^{(2)}&=& E_{\pi}^{(2)}(L,M)-E_0(L,M)
\end{eqnarray}
where $E_k(L,M)$ and $E_k^{(2)}(L,M)$ respectively denote
the lowest- and the second lowest eigenvalues of the Hamiltonian 
${\cal H}_0$ with the system size $L$, 
$\langle \sum _j (S^z_{1,j}+S^z_{2,j})\rangle=M$.
These gaps $\Delta_1$, $\Delta_\pi$ and $\Delta_\pi^{(2)}$
govern the long-distance behaviors the
$S^x-S^x$, $S^z-S^z$ and dimer correlations, respectively,
in the language of the pseudo-spin $\tilde{\bi S}$.
The critical indices $\eta$, $\eta^z$ and $\eta^{\rm d}$
can be connected with $\Delta_1$
and $\Delta_\pi$ by
\begin{equation}
    \eta = {L\Delta_1 \over \pi v_{\rm s}}~~~~~~
    \eta^z = {L\Delta_\pi \over \pi v_{\rm s}}~~~~~~
    \eta^{\rm d} = {L\Delta_\pi^{(2)} \over \pi v_{\rm s}}~~~~~~(L \to \infty)
\end{equation}
respectively,
where $v_{\rm s}$ is the spin-wave velocity estimated by
\begin{eqnarray}
  \label{vs}
  v_s={L \over {2\pi}}\left(E_{k_1}(L,M)-E_0(L,M)\right)~~~~~~
  (L \to \infty)
\end{eqnarray}
with $k_1 \equiv 2\pi/L$.

The behaviors of $\eta$ and $\eta^z$ near the KT critical point
for large $L$ are
\begin{equation}
    \eta = 1 - {1 \over 2}y_0~~~~~
    \eta^z = 1 - {1 \over 2}y_0 (1+2t)
    \label{eq:eta-y0}
\end{equation}
respectively,
where $t$ denotes the distance from the KT critical point,
and $y_0$ represents the lowest order finite-size correction, 
which leads to the logarithmic size correction
of the order of $1/\log L$ at the KT critical point.
Thus the KT critical point can be obtained from $\eta = \eta^z$ 
within the lowest order of the finite-size corrections.
As shown in equation (\ref{eq:eta-y0}),
the logarithmic corrections to $\eta$ and $\eta^z$ are serious
for finite systems.
Then the relation $\eta=1$ (or $\eta^z =1$) is not a good indicator
for the KT critical point.   

%
%  Fig. 3
%
\begin{figure}[h]
   \begin{center}
         \scalebox{0.5}[0.5]{\includegraphics{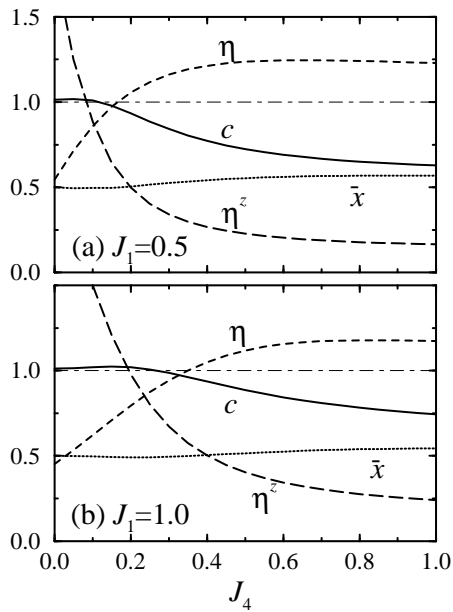}}
   \end{center}
   \caption{Critical exponents $\eta$ and $\eta^z$ as functions of 
            $J_4$ for (a) $J_1 = 0.5$ and (b) $J_1 = 1.0$ cases.
            The central charge $c$ (see equation (\ref{eq:c}))
            and the quantity $\overline{x}$ (see equation (\ref{eq:avex}))
            are also shown.}
   \label{fig:exp3}
\end{figure}

Figure 3 shows the behaviors of $\eta$ and $\eta^z$ as    
functions of $J_{4}$ for two cases 
(a) $J_1 = 0.5$ and (b) $J_1 = 1.0$ for the length of the 
ladder $L=16$.   
From the crossing points between $\eta$ and $\eta^z$,
we see that the critical point $J_{4\rm c} = 0.10$ 
for $J_1 = 0.5$, and $J_{4\rm c} = 0.22$ for $J_1 = 1.0$.
The quantity $1 - \eta = 1-\eta^z$ at $J_{4\rm c}$ represents
the magnitude of the logarithmic size correction $y_0/2$ around $L=8 \sim 16$. 
If we determine the KT critical points from the condition $\eta=1$,
we obtain $J_{4\rm c} = 0.18$ for $J_1 = 0.5$,
and $J_{4\rm c} = 0.36$ for $J_1 = 1.0$,
which are fairly larger than those by use of the LS.

To check the KT universality class,
we also calculated the central charge $c$ by use of
\begin{equation}
    \label{eq:c}
    {E_0(L,M) \over L}
    = \epsilon_{\rm g} 
      - {\pi cv \over 6L^2}
        \left\{ 1 + O\left({1 \over (\log L)^3} \right)\right\}
\end{equation}
where $\epsilon_{\rm g}$ is the ground-state energy per one rung 
at $m=1/2$ for an infinite system.    
As can be seen from figure \ref{fig:exp3},  
the central charge is $c=1$ near the KT critical point
and rapidly decreases in the gapped region, as is 
expected from \cite{inoue1,inoue2}.
This fact supports the KT nature of this transition.

We also checked the KT nature by calculating
\begin{equation}
    \label{eq:avex}
    \overline{x}
    = {(\eta^z + \eta^{\rm d})\eta \over 2}
\end{equation}
which should be $1/2$ in the lowest order of the renormalization
equation in the gapless region,  
as was shown by Kitazawa and Nomura \cite{kitazawa-nomura}.
We can clearly read from figure \ref{fig:exp3} that   
the quantity $\overline{x}$ is very close to $1/2$
in the gapless region, also verifying the KT nature.

%********
%** Fig.4
%*********
\begin{figure}[h]
\begin{center}
\scalebox{0.4}[0.4]{\includegraphics{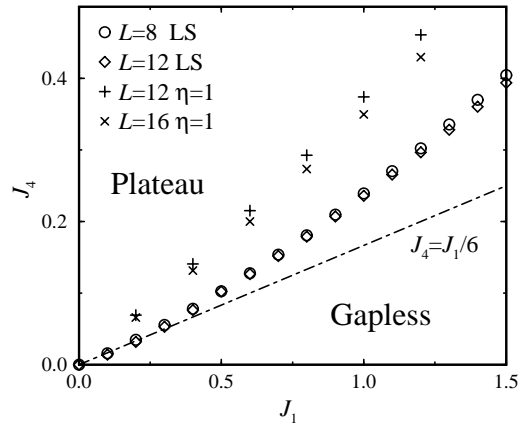}}
\end{center}
\caption{Phase diagram in the magnetized ground state at $m=1/2$. 
         Circles and diamonds are the critical points determined by LS 
         for $L=8$ and $L=12$, respectively. 
         They are almost independent of the system size. 
         The marks $+$ and $\times$ indicate the points where $\eta =1$ for 
         $L=12$ and $L=16$, respectively. They significantly depend on $L$.  
         The result from the strong rung coupling approach $J_{4c}\sim J_1/6$ 
         is also displayed. }
\label{fig:phase}
\end{figure}

In figure \ref{fig:phase} we show 
the KT critical points estimated by the LS (i.e. $\eta = \eta^z$),
as well as those obtained by assuming $\eta=1$ (conventional method), 
for the system sizes $L=8,12,16$. 
We see that the LS result is consistent with that by strong coupling
approach in \S 2, 
while the result obtained by the relation $\eta=1$ is not.

\section{Discussion}

In the previous section, 
we have obtained the phase diagram from the numerical diagonalization data
through the LS analysis, 
and shown the KT nature of the transition between the plateau and  
gapless states.  
Our numerical results are quite consistent with those of 
the strong coupling approach in \S3.  

In our previous work \cite{sh},
where we restricted ourselves to the case of $J_1 = 1$,
the KT transition was also suggested 
by the the Roomany-Wyld
approximation for the Callen-Symanzik $\beta$-function \cite{rw}. 
However, the critical exponent $\eta$ calculated numerically  
on the basis of the conformal field theory indicated 
a different value $\eta \sim 0.5$ around $J_{4c} \simeq 0.05$ estimated 
by the phenomenological renormalization.  
Since, as already shown in \$3,
the critical exponent $\eta$ should be unity at the KT transition point,
our previous result was not self-consistent.
This inconsistency has been completely solved by the present analysis. 
Namely, our previous result $J_{4c}=0.05 \pm 0.04$ for $J_1=1$ 
significantly deviates from the reliable result obtained above.  
It is consistent with the criticism \cite{level-1,level-2,nomura} that the 
phenomenological renormalization leads to serious 
underestimation of the 
gapless region for the KT transition. 

The phase diagram suggests that, for larger $J_1$,
the critical value $J_{4\rm c}$ becomes larger than 
the limit of a strong rung coupling.  
It implies that the quantum fluctuation due to the leg coupling ($J_{1}$) 
suppresses the plateau formation. 
In the special case $J_1=1$, the critical value is $J_{4c}\simeq 0.24$. 
Since Sr$_{14}$Cu$_{24}$O$_{41}$ is expected to have $J_1=1$ 
and $J_4 \sim 0.14$ as explained in \S1, 
it is not enough for the appearance of the plateau at $m=1/2$. 

An example of the magnetization curve was already shown in
our previous work \cite{sh}, to which readers should refer.  
At finite temperatures the magnetization plateau will be smeared,
which remains for future investigations.

In conclusion,  
we physically explained the plateau formation scenario
of the $S=1/2$ spin ladder with the ring exchange interaction at each 
plaquette, 
and also obtained the plateau-gapless phase diagram by
both analytical- and numerical methods.
The present phase diagram is more reliable and 
covers a wider range of parameters
than our previous one \cite{sh}.

\bigskip
\section*{Acknowledgement}

We wish to thank Koyohide Nomura, Yasushi Honda, 
Takeshi Horiguchi for fruitful
discussions.
We also thank
the Supercomputer Center, Institute for
Solid State Physics, University of Tokyo for computer facilities. 
This work was 
supported in part by Grant-in-Aid for the Scientific Research Fund
from the Ministry of Education, Science, Sports and Culture (11440103).

\end{document}